# Sojourn-time distribution of virus capsid in interchromatin corrals of a cell nucleus

YUICHI ITTO [a] and JENS B. BOSSE [b]

[a] Science Division, Center for General Education, Aichi Institute of Technology,
Aichi 470-0392, Japan
[b] Heinrich Pette Institute, Leibniz Institute for Experimental Virology,
Hamburg, Germany

**Abstract** Virus capsids in interchromatin corrals of a cell nucleus are experimentally known to exhibit anomalous diffusion as well as normal diffusion, leading to the Gaussian distribution of the diffusion-exponent fluctuations over the corrals. Here, the sojourn-time distribution of the virus capsid in local areas of the corral, *i.e.*, probability distribution of the sojourn time characterizing diffusion in the local areas, is examined. Such an area is regarded as a virtual cubic block, the diffusion property in which is normal or anomalous. The distribution, in which the Gaussian fluctuation is incorporated, is shown to tend to slowly decay. Then, the block-size dependence of average sojourn time is discussed. A comment is also made on (non-)Markovianity of the process of moving through the blocks.

# 1. Introduction

Viruses exhibit rich phenomena, which are highly attractive from the viewpoint of physics [1-3]. In particular, a recent experimental study [4] (see also Ref. [5]) has reported a remarkable finding for the diffusion phenomenon of herpesviruses in nuclei of PtK2 cells. There, the cells were infected with pseudorabies virus (*i.e.*, suid herpesvirus 1) or herpes simplex virus 1. Capsid, which is a protein shell surrounding viral DNA, has been labeled with a fluorescent protein, and tracks of such virus capsids have been observed in the nucleus by the single particle tracking. The experiment has shown that the virus capsids diffuse in interchromatin compartments, which are formed by chromatin (*i.e.*, chromosomal substance). Such a compartment is called corral. It has also been found that, during virus infection, chromatin structure becomes more porous, which makes the corral size increase. As a result, the virus capsid moves through the corrals in order to reach nuclear membrane.

The diffusion property has been characterized by the mean square displacement of the virus capsid based on each track:

$$\overline{x^2} \sim t^\alpha, \tag{1}$$

where $t$ is elapsed time and $\alpha$ is termed the diffusion exponent. Then, the diffusion exponent fluctuates depending on the corrals in a wide range of $\alpha$ [4,5]. More precisely, the exponent takes not only $\alpha = 1$, *i.e.*, normal diffusion, but also $\alpha \neq 1$, *i.e.*, anomalous diffusion: subdiffusion (superdiffusion) in the case with $0 < \alpha < 1$ ($\alpha > 1$). Then, $\alpha$ obeys the following Gaussian distribution:

$$f(\alpha) \sim \exp\left[-\frac{(\alpha - \alpha_0)^2}{2\sigma^2}\right], \qquad (2)$$

where $\alpha_0 = 0.85$ and $\sigma = 0.24$ are, respectively, the mean value and the standard deviation of $\alpha$. Thus, the diffusion observed offers an outstanding feature in anomalous diffusion [6-8] under vital investigation in the literature.

Regarding the Gaussian fluctuation in Eq. (2), a fact to be emphasized is its *robustness* in the sense [4,5] that it takes the same form for both types of the virus (*i.e.*, the pseudorabies virus and the herpes simplex virus 1). Thus, this is seen to manifest the existence of universality of the fluctuations over the corrals. We point out that the Gaussian distribution in Eq. (2) has theoretically been derived in a consistent manner in a recent work [9].

In such a situation, a central issue is to understand fundamental dynamics of the virus capsid in the corrals. In fact, to clarify its spatial property, the distribution of the spatial displacement of the capsid in the corrals has been analyzed [4]. The analysis is seen to suggest that this distribution may obey an exponential law, which means that large displacements are not significant. In the random walk scheme [10], this may imply [11] that the spatial property is trivial for the origin of subdiffusion as well as normal diffusion but is not in the case of superdiffusion, where the presence of such displacements is necessary like, *e.g.*, Lévy flights [12], showing how exotic the dynamics of the virus capsid is.

In the present work, we focus our attention on the *temporal* property of diffusion of the virus capsid in the corrals. In particular, we examine the sojourn-time distribution, which is meant as probability distribution of the sojourn time characterizing diffusion in local areas of the corral. Such local areas are treated as virtual cubic blocks, in which the virus capsid exhibits normal or anomalous diffusion. We show that the distribution, in which the Gaussian fluctuation in Eq. (2) is incorporated, tends to slowly decay. Then, we analyze the dependence of average sojourn time on the block size based on the experimental data. In addition, we present a brief discussion about (non-)Markovianity of the process of moving from one block to another in connection with a relation to be satisfied by a class of Markovian processes. This may offer a possible way of examining if there exists a long-term memory in the process or not. The present work is expected to contribute to deeper understanding diffusion of the virus capsid in the nucleus.

Throughout the present work, variation of the fluctuations is considered to be very slow on a long time scale, on which the Gaussian fluctuation in Eq. (2) is realized. In other words, $\alpha$ is supposed to be approximately constant on such a time scale. Then, the minimum value of $\alpha$ is not zero and $f(\alpha) \sim 0$ for such a value, which is supported by the raw data of the experiment [4].

## 2. Sojourn-time distribution

Consider diffusion of the virus capsid over the region of the corrals in the nucleus. Like in recent works [13,14], we regard this region as a medium for diffusion of the capsid at both viral types, which is composed of many virtual cubic blocks. The virus capsid exhibits normal or anomalous diffusion, depending on these local blocks. It should be noticed [4] that $\overline{x^2}$ in Eq. (1) has been analyzed for the elapsed time smaller than that for determination of the corral size. Accordingly, the block is identified with a local area of the corral. To evaluate the block size, we express Eq. (1) more precisely as $\overline{x^2} = 6Dt^\alpha$ [4]. $D$ is a generalized diffusion coefficient denoted here as $D = \Delta^2 / s^\alpha$, where $\Delta$ and $s$ are positive characteristic constants describing a spatial displacement of the virus capsid and time being required for the displacement, respectively. Then, let $l \equiv \sqrt{\overline{x^2}}$ be the length of the side of the cubic block. Thus, the block size, $l$, is determined in this way.

As can be seen from the above discussion, $l$ depends on the elapsed time, given a value of $\alpha$. This, in turn, means that the elapsed time depends on $\alpha$, given a value of $l$, which is given by

$$t = \left( \frac{l}{\sqrt{6}\,\Delta} \right)^{2/\alpha} s. \qquad (3)$$

This offers the *sojourn time* characterizing diffusion of the virus capsid in a given local block. Then, what we are interested in here is its statistical property, in which the Gaussian fluctuation in Eq. (2) is taken into account. This shows how long/short the virus capsid stays in a given local block at the statistical level. The virus capsid moves through the local blocks with different exponents over the medium and therefore the block size can have different values. However, in order to clarify the statistical property as simple as possible, in the present paper, we discuss the case when $l$ is set as a certain value for all of the blocks in the medium. In other words, the medium is divided into the blocks with an equal block size. [Note that $\alpha$ fluctuates according to the distribution in Eq. (2).] Thus, in such a case, $t$ in Eq. (3) becomes a random variable, $t = t(\alpha)$: the randomness comes from the diffusion exponent, $\alpha$. It is noticed that we are interested in the case when $t > s$, which requires the block size to fulfill $l > \sqrt{6}\Delta$.

We denote the probability of finding a certain value of $t$ in the interval $[\tau, \tau+d\tau]$ by $P(\tau)d\tau$. This describes the sojourn-time distribution, which can be formally given as follows:

$$P(\tau) = \langle \delta(\tau - t(\alpha)) \rangle_\alpha \qquad (4)$$

with $\langle \bullet \rangle_\alpha$ being the average over the Gaussian distribution in Eq. (2). Therefore, in the case when $\tau > s$, substitution of Eq. (3) into Eq. (4) leads to the following form of the sojourn-time distribution:

$$P(\tau) \sim \frac{1}{\tau[\ln(\tau/s)]^2} \exp\left\{-\left[2\frac{\ln[l/(\sqrt{6}\Delta)]}{\ln(\tau/s)} - \alpha_0\right]^2 /(2\sigma^2)\right\}, \qquad (5)$$

showing that the distribution tends to slowly decay, since it is logarithmically related to $\tau$.

### 3. Block-size dependence

As mentioned earlier, it has been observed [4] that the corral size increases, since chromatin structure becomes more porous during virus infection. This seems to correspond to the situation that the block size increases. So, it may be of interest to evaluate the block-size dependence of average sojourn time, which is given, from Eq. (4), by the average of $t(\alpha)$ with respect to $f(\alpha)$: $\langle t(\alpha) \rangle_\alpha = \langle t(\alpha) \rangle_\alpha (l)$. Below, we wish to discuss this issue based on the experimental data.

Here, the characteristic constants and the range of the block size are taken as follows. According to the experiment [4], to determine the positions of the virus capsids at both viral types, the cells have been imaged at a frame rate of 36 frames per second, for which the capsids are present in each consecutive frame. This seems allow us to choose the value of $s$ as the inverse of the rate, $s = 0.028\,\text{s},$ for both viral types. Then, it has also been shown that $D = 0.035\,\mu\text{m}^2/\text{s}^\alpha$ with $\alpha = 0.961$ for the pseudorabies virus, whereas $D = 0.023\,\mu\text{m}^2/\text{s}^\alpha$ with $\alpha = 0.918$ for the herpes simplex virus 1. Although these values are obtained from average over not a given track but all of the tracks, we employ them as the representative values of both $D$ and $\alpha$. Therefore, the values of $\Delta$ are estimated as follows: $\Delta = 0.034\,\mu\text{m}$ for the pseudorabies virus, whereas $\Delta = 0.029\,\mu\text{m}$ for the herpes simplex virus 1. Regarding the value of $l$, it should be larger than $0.125\,\mu\text{m},$ which is the diameter of the virus capsid as a sphere. Then, we suppose that the block size has its largest value given by $l = \sqrt{\overline{x^2}}$ with the above representative values at $t = 0.36\,\text{s},$ since the Gaussian distribution in Eq. (2) has been observed at such an elapsed time [4]. Such largest values are estimated as $0.28\,\mu\text{m}$ and $0.23\,\mu\text{m}$ for the pseudorabies virus and the herpes simplex virus 1, respectively.

In Fig. 1, we present the plots of $\langle t(\alpha) \rangle_\alpha$ for both viral types, in which the raw data of $\alpha$ obtained in the experiment [4] has been employed. There, we see that $\langle t(\alpha) \rangle_\alpha$ monotonically increases with respect to $l$ at both types. In addition, the value of $\langle t(\alpha) \rangle_\alpha$ for the pseudorabies virus is seen to be smaller than that for the herpes simplex virus 1 at each value of $l$, indicating that the capsid of the former diffuses faster than that of the latter.

## 4. Comment on (non-)Markovian nature

The virus capsid moves from one cubic block to another. In other words, the virus capsid passes through the boundary of a given cubic block. We treat this as an event. In this section, we make a comment on (non-)Markovianity of the process of such events over the medium. For it, we examine a relation to be satisfied by a class of Markovian processes, which has been discussed in the context of laser cooling of atoms in Ref. [15].

Suppose a sequence of the events occurred in the time interval $[0, t]$, where the same symbol $t$ as that in Eq. (1) is used for time (*i.e.*, the conventional time) but it will not cause confusion. In this situation, $P(\tau)$ in Eq. (4) seems to play a role of the distribution of the time interval between two successive events. Let us denote the mean density of events at time $t$ by $S(t)$. If the process is Markovian, then the following equation holds [15]:

$$S(t) = P(t) + \int_0^t dt' P(t-t') S(t'). \qquad (6)$$

Since this includes a convolution integral, the Laplace transformation of Eq. (6) yields

$$\mathcal{L}[S](u) = \frac{\mathcal{L}[P](u)}{1 - \mathcal{L}[P](u)}, \qquad (7)$$

where $\mathcal{L}[g](u) \equiv \int_0^\infty dt\, e^{-ut} g(t)$. We are particularly interested in the long time behavior of $S(t)$. Using Eq. (4), $\mathcal{L}[P](u)$ is therefore calculated, up to the first order of $u$ (*i.e.*, small-$u$ behavior), to be

$$\mathcal{L}[P](u) \sim 1 - \langle t(\alpha) \rangle_\alpha u, \tag{8}$$

leading to

$$\mathcal{L}[S](u) \sim \frac{1}{\langle t(\alpha) \rangle_\alpha} \frac{1}{u}. \tag{9}$$

Thus, as the long time behavior, we obtain

$$S(t) \sim \frac{1}{\langle t(\alpha) \rangle_\alpha}. \tag{10}$$

Accordingly, the mean density of events behaves as a certain constant at the block size under consideration, which is equal to the inverse of average sojourn time. In this respect, the analysis performed on $\langle t(\alpha) \rangle_\alpha$ in the previous section tells us about the dependence of the mean density on the block size.

The above observation means that, if the process is Markovian, then the virus capsid moves through the blocks at the rate of $1/\langle t(\alpha) \rangle_\alpha$ for large time. Although such a feature is quite general for distributions with finite first moment, the point is in clarifying if the mean density to be observed deviates from the above rate or not. In other words, the violation of Eq. (10) implies the presence of a long-term memory in the process, indicating that the events are not temporally separable. Then, this may be the case, as we shall see below.

In Fig. 5 (B) in Ref. [4], the probability distribution of the spatial displacement of the capsid over $1.5\text{s}$ has been presented in the case of the pseudorabies virus. So, we here consider that a typical scale of the displacement, $\Delta^*$, is given by the spatial extension of the distribution such as its half-width: $\Delta^* \sim 0.05\mu\text{m}$. It seems therefore that the number of events in a certain time subinterval is proportional to $\Delta^*/(1.5l)\,[1/\text{s}]$. Now, as can be seen from Fig. 1, this quantity deviates from $1/\langle t(\alpha)\rangle_\alpha$ at each value of $l$ in the same subinterval. For example, the ratio of the former to the latter at $l = 0.28\,\mu\text{m}$ is about 0.67. Thus, these facts may imply a possible violation of Eq. (10) and accordingly may support non-Markovianity of the process.

We note, however, the following points. In the present case, the sojourn time in the local block is regarded as the time interval between two successive events. Therefore, further studies based on the time interval taken from a set of the time series data of the process seem to be needed for examining (non-)Markovianity of the process.

## 5. Concluding remarks

We have examined the sojourn-time distribution of virus capsid diffusing in interchromatin corrals over nucleus of PtK2 cell for pseudorabies virus and herpes simplex virus 1. We have regarded the region of the corrals as a medium consisting of virtual cubic blocks with equal size. Taking the Gaussian fluctuation of the diffusion exponent into account, we have shown that the distribution tends to slowly decay. Combined with the raw data of experiment, we have performed the analysis of the block-size dependence of average sojourn time. We have also made a comment on (non-)Markovian nature of the process of moving through the blocks.

We point out the following. As mentioned in the Introduction, the dynamics of the virus capsid in the corrals seems to be exotic. A recent experimental work in Ref. [16] may provide further information about this point, in which simulation analysis based on the random walk scheme has been performed for herpesvirus capsids. In addition, it may be of interest to examine application of a theoretical framework in Ref. [17] (see also Ref. [18]) based on the fluctuation distribution for describing diffusion of the virus capsid over the corrals.

**Acknowledgements**

We thank Lynn Enquist (Princeton University) for support. Y.I. would like to acknowledge the Heinrich Pette Institute, Leibniz Institute for Experimental Virology for the nice working atmosphere, which enabled the present work. He also thanks S. Abe for informative discussions. He was supported by the foreign research project of the Aichi Institute of Technology. The Heinrich Pette Institute, Leibniz Institute for Experimental Virology is supported by the Freie und Hansestadt Hamburg and the Bundesministerium für Gesundheit (BMG).

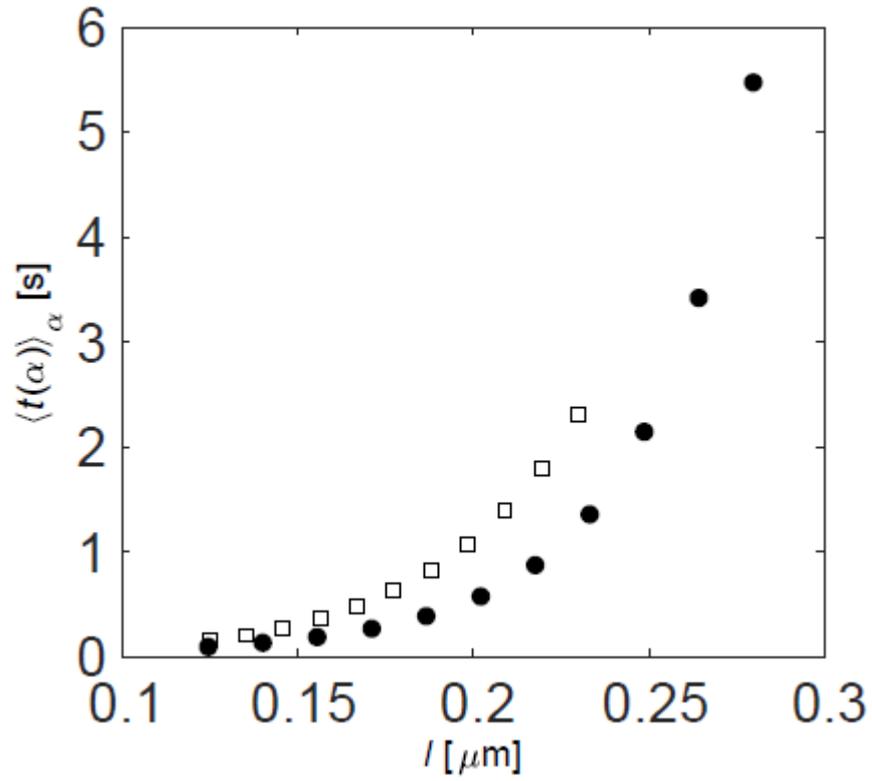

**Fig. 1** The block-size dependence of average sojourn time, $\langle t(\alpha) \rangle_\alpha$, of the virus capsids. The filled circles and open squares are for the pseudorabies virus and the herpes simplex virus 1, respectively.